# A new topological semimetal with iso-energetic Weyl fermions in TaAs under high pressure

Yonghui Zhou[1,2*], Pengchao Lu[3*], Yongping Du[3*], Xiangde Zhu[1*], Ranran Zhang[1], Dexi Shao[3], Xuliang Chen[1,2], Xuefei Wang[2], Mingliang Tian[1,4], Jian Sun[3,4†], Xiangang Wan[3,4†], Zhaorong Yang[1,2,4†], Yuheng Zhang[1,4] & Dingyu Xing[3,4]

[1] High Magnetic Field Laboratory, Chinese Academy of Sciences, Hefei 230031, China.

[2] Key Laboratory of Materials Physics, Institute of Solid State Physics, Chinese Academy of Sciences, Hefei 230031, China.

[3] National Laboratory of Solid State Microstructures, College of Physics, Nanjing University, Nanjing 210093, China.

[4] Collaborative Innovation Center of Advanced Microstructures, Nanjing University, Nanjing 210093, China.

*These authors contributed equally to this work.

†Correspondence and requests for materials should be addressed to J.S. (e-mail: jiansun@nju.edu.cn); or to X.W. (e-mail: xgwan@nju.edu.cn); or to Z.Y. (e-mail: zryang@issp.ac.cn).

TaAs as one of the experimentally discovered topological Weyl semimetal has attracted intense interests recently. The ambient TaAs has two types of Weyl nodes which are not on the same energy level. As an effective way to tune lattice parameters and electronic interactions, high pressure is becoming a significant tool to explore new materials as well as their exotic states. Therefore, it is highly interesting to investigate the behaviors of topological Weyl fermions and possible structural phase transitions in TaAs under pressure. Here, with a combination of *ab initio* calculations and crystal structure prediction techniques, a new hexagonal *P*-6*m*2 phase is predicted in TaAs at pressure around 14 GPa. Surprisingly, this new phase is a topological semimetal with only single set of Weyl nodes exactly on the same energy level. The phase transition pressure from the experimental measurements, including electrical transport measurements and Raman spectroscopy, agrees with our theoretical prediction reasonably. Moreover, the *P*-6*m*2 phase seems to be quenched recoverable to ambient pressure, which increases the possibilities of further study on the exotic behaviors of single set of Weyl fermions, such as the interplay between surface states and other properties.

# Introduction

Weyl semimetals (WSMs) can be considered as three-dimensional (3D) analogs to graphene in terms of the electronic dispersions. One of the most striking characteristics of a WSM is the topological surface state with Fermi arcs, which is originally proposed in pyrochlore iridates[1], and further studied in different systems by other groups[2,3,4,5], while the bulk may possess Weyl nodes behaving like Weyl fermions[1,2] and novel transport properties stemming from the chiral anomaly[6,7,8]. Weyl nodes appear in pairs of opposite chirality due to the 'No-go theorem'. Recently, the non-centrosymmetric NbAs-type transition-metal monoarsenides, i.e. TaAs, TaP, NbAs, and NbP, have been predicted to be WSMs, and twelve pairs of Weyl nodes are expected in their 3D Brillouin zones[9,10]. Moreover, NbAs-type transition-metal monoarsenides are completely stoichiometric and nonmagnetic, providing a proper platform for the study of the chiral anomaly in the topological WSMs. Soon after, many exotic properties induced by the Weyl nodes have been observed such as topological surface states with Fermi arcs[11,12,13,14] and a negative longitudinal magnetoresistance (MR)[15,16,17] due to the chiral anomaly[18,19,20,21,22]. Most remarkably, bulk-sensitive angle-resolved photoemission spectroscopy (ARPES) measurements[13,23] have shown Weyl nodes in the bulk states of TaAs in good agreement with *ab initio* calculations[9,10,13]. These results demonstrate that TaAs family is a promising system that hosts topological properties and might help to pave a new way for further experimental studies of new topological phases and quantum states.

However, there are two types of Weyl nodes on different energy levels in TaAs, which makes the observation of Fermi arcs quite complicated[10,11]. In that sense, it might be useful to reduce the interference from the bulk states, if one can find a material with only one set of Weyl points on the same energy level. On the other hand, as one of the fundamental state parameters, high pressure is an effective and clean way to tune lattice as well as electronic states, especially in quantum states[24,25,26]. Therefore, it will be very interesting and important to investigate whether these two sets of Weyl points in TaAs would move or merge into one level in energy under high pressure, or whether pressure can induce some other exotic structures and properties in these topological WSMs.

In this work, we report a joint study on the pressure effects on the electronic and structural properties of single-crystal TaAs, from both *ab initio* calculations and experimental measurements, including electrical resistance measurements and Raman spectroscopy. Our numerical simulations demonstrate that the Weyl nodes in the *I*4$_1$*md* structure remain stable upon compression. Most interestingly, using crystal structure searching techniques and *ab initio* calculations, a new *P*-6*m*2 phase with only one set of Weyl nodes is predicted above 14 GPa. Furthermore, the critical transition pressures from electrical resistance and Raman measurements both agree reasonably with theoretical predictions.

## Method

***Ab initio* calculations.** We use *ab-initio* random structure searching (AIRSS)[27,28]

method for crystal structure predictions. Cell optimization are performed using projector augmented wave (PAW) implemented in the Vienna *ab initio* simulation package (VASP)[29], with Perdew-Burke-Ernzerhof (PBE) generalized gradient approximation (GGA) exchange-correlation density functional[30]. Phonon spectra are carried out using finite displacement method with VASP and PHONOPY[31] code together. The plane wave cutoff is 400 eV, and the Brillouin zones are sampled with a Monkhorst-Pack **k**-mesh with a spacing of 0.03 Å$^{-1}$. A $3 \times 3 \times 3$ supercell for *P*-6*m*2 and a $2 \times 2 \times 2$ supercell for *P*2$_1$/*c* phase, are chosen for phonon calculations. Electronic structure calculations are performed using the full-potential linearized augmented plane-wave (FP-LAPW) method implemented in the WIEN2k package[32]. A 1000 **k**-point mesh for BZ sampling and -7 for the plane wave cut-off parameter $R_{MT}K_{max}$ are used in the calculation, where R is the minimum LAPW sphere radius and $K_{max}$ is the plane-wave vector cutoff. Spin-orbit coupling is taken into account by a second-variation method[33].

**High pressure experiments.** Single crystals of TaAs used here were grown by chemical vapor transport method, and the crystal structure and quality were characterized by X-ray diffraction and energy dispersive X-ray spectroscopy (EDXS) measurements. Electrical transport measurements at ambient pressure were performed in a Physical Property Measurement System (PPMS, Quantum Design). High-pressure resistance measurements were conducted via the standard four-probe method in a screw-pressure-type diamond anvil cell (DAC). Raman experiments were performed at room temperature in a BeCu-type Diacell ST-DAC using a Horiba Jobin

Yvon T64000 spectrometer. These measurements were conducted with the 514.5 nm excitations of an Ar-Kr laser with a power below 0.1 mW to avoid sample damage and any heating effect. Pressure was calibrated by using the ruby fluorescence shift at room temperature for all experiments[34].

## Results and discussions

The ambient TaAs crystallizes in a body-centered tetragonal NbAs-type structure with the non-centrosymmetric space group[35]. This structure consists of four trigonal prisms per unit cell, in which the alternating Ta and As layers are rotated by 90 ° and shifted by $a/2$, as shown in Fig. 1(c). The absence of a horizontal mirror plane in the unit cell thus breaks the inversion symmetry. It is known that the ambient $I4_1md$ structure has two sets of Weyl points. One type of Weyl points locates in the $k_z = 0$ plane (W1), and the other is off this plane (W2)[10,11]. These two sets of Weyl points have an energy difference about 14 meV at ambient pressure[11]. To check if pressure can eliminate this energy difference, we performed additional calculations and found that with the increasing pressure, as shown in Table I, the $x$ coordinate of W1 and $z$ coordinate of W2 decrease monotonously, while the rest coordinates of W1 and W2 are almost unchanged. Thus the Weyl points in $I4_1md$ structure are quite robust upon compression. This is good at one side, however on the other side, the energy difference between these two sets of Weyl points still remains and even becomes larger under moderate pressure. This encourages us to study whether higher pressure can eliminate this energy difference eventually and to look for the possible new structure with iso-energetic Weyl points.

To accomplish this, we used crystal structure prediction techniques in the frame of density functional theory (DFT) to find the best candidates of TaAs under high pressure. The enthalpy-pressure ($\Delta H$-$P$) curves plotted in Fig. 1(a) exhibit the best ones from our structural predictions under pressure up to 50 GPa. The ambient structure of TaAs is found to be $I4_1md$, which is in agreement with the experimental observations[35]. Moreover, the enthalpy of $P$-$6m2$ structure is found to be lower than that of the ambient $I4_1md$ structure from 14 GPa to 34 GPa. The calculated transition pressure from $I4_1md$ to $P$-$6m2$ is around 14 GPa. Interestingly, these two structures are formed with the same type $TaAs_6$ polyhedron, a triangular prism, of which triangle side length and height are essentially the same, respectively (3.305 Å and 3.396 Å for $I4_1md$, 3.300 Å and 3.456 Å for $P$-$6m2$ at 15 GPa). As shown in Fig. 1(c), the main difference to distinguish these two structures is the stacking sequence of the triangular prisms. While the $I4_1md$ structure is constructed with *AB* sequence along the *c*-axis, the $P$-$6m2$ structure has a uniform *AA* stacking. When the pressure increases, a $P2_1/c$ structure becomes the most stable one beyond 34 GPa. The crystal structure of the $P2_1/c$ phase is more complicated, which consists of twisted $TaAs_7$ enneahedra.

Calculated band structures and Fermi surfaces of the high-pressure phases are shown in Fig. 2. For the $P$-$6m2$ structure, it seems to be metallic as the bands are overlapping near the A point. Spin-orbit coupling opens a gap of around 160 meV for the crossing point near the K point. The density of states near the Fermi level is mainly contributed by the Ta-5$d$ orbits. Due to the effect of triangular prism crystal field, Ta-5$d$ orbits are splitted into three groups: $d_{z^2}$, $d_{xz}+d_{yz}$ and $d_{x^2y^2}+d_{xy}$. There are four bands contributing to the Fermi surface, and interestingly, one of them forms a doughnut surrounding the K point, and two balls in the Γ-A path near the A point, which is mainly from Ta $d_{z^2}$ and $d_{x^2y^2}+d_{xy}$ orbits. Meanwhile, the other three bands constitute a bowl surrounding the A point, contributed by Ta $d_{x^2y^2}+d_{xy}$ orbits.

The intriguing electronic band structures of *P*-6*m*2 encourage us to investigate its topological properties. Interestingly, our calculations show that this non-centrosymmetric structure is also a Weyl semimetal with only one type of Weyl points, which are around the K point. As shown in Fig. 2(c), there are 12 Weyl nodes in the first Brillouin Zone. The Weyl points obtained from scanning in the Γ-K-M and Γ-K-A planes are shown in Fig. 2 (d) and (e), respectively. The iso-energetic feature might be important for the possible application of Weyl semimetals in microelectronics. As presented in table II, the position of these Weyl nodes almost stay unchanged, which is different from the behaviors of W1 and W2 in the *I*4$_1$*md* phase.

For the *P*2$_1$/*c* structure, there are three bands across Fermi level, near the Z, Y, A, B, C point, mainly composed of Ta-5$d$ orbits, making some cobbles (two surrounding Z point and four near Y point) and twisted tubes connecting B, A, C points. All the five components of Ta-5$d$ orbits contribute together to these three bands on the Fermi level. As a nonmagnetic compound with inversion symmetry, no Weyl nodes are expected in this phase.

Phonon spectra of *P*-6*m*2 and *P*2$_1$/*c* structures are shown in supplementary Fig. S1. They are showed to be dynamically stable at 25 and 40 GPa. Moreover, the phonon spectra of these two structures at 0 GPa do not have modes with negative frequencies, which indicates *P*-6*m*2 and *P*2$_1$/*c* structures should be quenched recoverable to ambient pressure if they could be synthesized.

To study the influence to the transport properties from pressure, we perform detailed high pressure electrical resistance measurements. Supplementary Fig. S2(b) displays the temperature-dependent in-plane resistivity of single-crystal TaAs at ambient pressure. At zero magnetic field, TaAs exhibits a metallic-like behavior down to 2 K and a residual resistance ratio (RRR = 25). The applied magnetic field of 2 T perpendicular to the current and the *ab* plane significantly increases the resistivity below ~200 K and induces a crossover from metallic-like to insulating-like conductivity. As seen from supplementary Fig. S2(c,d), the Shubnikov-de Haas (SdH) oscillations and the Landau fan diagram suggest the presence of the π Berry's phase of the Weyl electron pocket[16,38]. All of these results are consistent with recent reports[15,16], which confirms the high quality of TaAs sample used here. After applying external pressure of 1.1 GPa, the temperature dependence of resistance exhibits an insulating-like behavior (*dρ/dT* < 0) over the whole measured temperature range, as shown in Fig. 3(a). The pressure-induced metal-insulator transition intuitively resembles the case of magnetic field effect on the resistance (see supplementary Fig. S2(b)). Nevertheless, this phenomenon is counterintuitive because pressure usually enhances the band overlapping. Note that similar phenomenon has been reported in pressurized 3D Dirac semimetal $Cd_3As_2$[39,40]. However, this is not the case of isostructural NbAs which remains metallic behavior up to 20 GPa[41]. In the pressure range from 1.1 to 10.0 GPa, the $R(T)$ curve keeps the insulating-like behavior. The resistance at 300 K increases monotonically, while the resistance at 1.8 K first increases with pressure then decreases at pressures above 5.0 GPa (see Fig. 3(c)). Accordingly, when the pressure is enhanced to 14.0 GPa, an insulating-like to metallic-like transition emerges upon cooling, showing a resistance anomaly at $T_{M-I}$ ~140 K. With further increasing pressure, the peak temperature $T_{M-I}$ shifts to higher temperatures (see Fig. 3(c)). The transition pressure agrees with the $I4_1md$ to $P$-6$m$2 transition from the theoretical prediction, as shown in Fig. 1(a).

Since TaAs possesses large MR in analogy to $WTe_2$[42], which has been predicted very recently to be a new type of WSMs[43], we investigated the MR at 4.5 K under different pressures. As seen in Fig. 3(d), the MR is dramatically suppressed by pressure, which is also similar to the case of pressurized $WTe_2$[44]. In $WTe_2$, the suppression of MR is accompanied by the emergence of superconductivity[44,45]. However, in our case, no trace of resistance drop appears even down to 1.8 K till 54.0 GPa. This result somewhat resembles that of NbAs, in which no dramatic pressure effect was observed on both crystal structure and electronic state up to 20.0 GPa[41]. Nevertheless, it is interesting to note that the MR curve at 10.0 GPa displays a linear behavior in a broad field region. The linear MR seems to be universal in the Dirac or Weyl semimetals, strongly suggesting the robust Weyl semimetal state in TaAs even under high pressure. Our calculations also show that the Weyl states in $I4_1md$ remains stable up to 14.0 GPa. As shown in Fig. 3(d), the characteristic field $B^*$ above which the linear MR is observed first shifts monotonically to lower fields with increasing pressure from 3.0 GPa, then turns back for pressures above 10.0 GPa. Combining with the fact that an insulating-like to metallic-like transition emerges upon cooling above 10.0 GPa, it can be deduced that this pressure is close to a critical pressure corresponding to a possible structural or topological phase transition. The critical pressure close to 10.0 GPa is also revealed in Fig. 3(c), where the room-temperature resistance displays a maximum at 14.0 GPa, levels off in the pressure region from 20.0 to 37.0 GPa, and then drops slightly afterwards. Unlikely to be coincidence, these transition points sit more or less in the same pressure range for the occurrence of the high-pressure phases $P$-$6m2$ and $P2_1/c$, as predicted in theoretical calculations.

To have a better understanding of the phase transition of TaAs, we measured high-pressure Raman scattering spectroscopy, which is sensitive to local bond vibrations and symmetric broken. The non-centrosymmetric structure of TaAs is characterized by the space group $I4_1md$ ($D_{4v}^{11}$, No.109), with four atoms in unit cell, where the Wyckoff positions are $4a$ for both Ta atoms and As atoms[35]. As shown in Fig. 4 in the wavenumber range from 200 and 400 cm$^{-1}$, there is one vibration mode near 252.7 cm$^{-1}$ at ambient pressure, which is designated to $A_1$ mode and compatible with the reported results observed in the $x'x'$ spectra[46]. After applying pressure, this mode shifts toward higher frequency. When the pressure is enhanced to 10.1 GPa which is close to the critical pressure as revealed in the transport measurements, this mode cannot be detected within the system resolution. This is probably the Raman spectroscopic evidence for structural phase transition, suggesting the intimate correlation between the lattice structure and metal-insulator transition in TaAs under high pressure. No modes can be observed as the pressure is further increased up to 27.2 GPa.

# Conclusion

In conclusion, by combining theoretical and experimental investigations, we find a new hexagonal *P*-6*m*2 phase in TaAs at around 14 GPa. This structure is predicted to be a topological semimetal with single set of Weyl fermions. This property might be good to reduce the interference between the surface and bulk states. In addition, we find the Weyl states in the ambient tetragonal *I*4$_1$*md* structure remain stable with pressure up to 14 GPa, this somehow agrees with the linear MR observed in the transport measurements at 10 GPa. The behaviors of iso-energetic Weyl fermions in this material might provide an important platform to study the interplay between surface states and other exotic properties and desire further investigations.

# Acknowledgements

This research was financially supported by the National Key Projects for Basic Research of China (Grant Nos: 2015CB921202), the National Natural Science Foundation of China (Grant Nos: 51372112 and 11574133), NSF Jiangsu province (No. BK20150012), and the Fundamental Research Funds for the Central Universities. Part of the calculations was performed on the supercomputer in the High Performance Computing Center of Nanjing University.

# References

1 Wan, X., Turner, A. M., Vishwanath, A. & Savrasov, S. Y. Topological semimetal and Fermi-arc surface states in the electronic structure of pyrochlore iridates. *Phys. Rev. B* **83**, 205101, doi:10.1103/PhysRevB.83.205101 (2011).

2 Xu, G., Weng, H., Wang, Z., Dai, X. & Fang, Z. Chern Semimetal and the Quantized Anomalous Hall Effect in $HgCr_2Se_4$. *Phys. Rev. Lett.* **107**, 186806, doi:10.1103/PhysRevLett.107.186806 (2011).

3 Hosur, P. Friedel oscillations due to Fermi arcs in Weyl semimetals. *Phys. Rev. B* **86**, 195102, doi:10.1103/PhysRevB.86.195102 (2012).

4 Ojanen, T. Helical Fermi arcs and surface states in time-reversal invariant Weyl semimetals. *Phys. Rev. B* **87**, 245112, doi:10.1103/PhysRevB.87.245112 (2013).

5 Potter, A. C., Kimchi, I. & Vishwanath, A. Quantum oscillations from surface Fermi arcs in Weyl and Dirac semimetals. *Nat. Commun.* **5**, 5161, doi:10.1038/ncomms6161 (2014).

6 Zyuzin, A. A. & Burkov, A. A. Topological response in Weyl semimetals and the chiral anomaly. *Phys. Rev. B* **86**, 115133, doi:10.1103/PhysRevB.86.115133 (2012).

7 Wang, Z. & Zhang, S.-C. Chiral anomaly, charge density waves, and axion strings from Weyl semimetals. *Phys. Rev. B* **87**, 161107(R), doi:10.1103/PhysRevB.87.161107 (2013).

8 Parameswaran, S. A., Grover, T., Abanin, D. A., Pesin, D. A. & Vishwanath, A. Probing the Chiral Anomaly with Nonlocal Transport in Three-Dimensional Topological Semimetals. *Phys. Rev. X* **4**, 031035, doi:10.1103/PhysRevX.4.031035 (2014).

9 Huang, S.-M. *et al.* A Weyl Fermion semimetal with surface Fermi arcs in the transition metal monopnictide TaAs class. *Nat. Commun.* **6**, 7373, doi:10.1038/ncomms8373 (2015).

10 Weng, H., Fang, C., Fang, Z., Bernevig, B. A. & Dai, X. Weyl Semimetal Phase in Noncentrosymmetric Transition-Metal Monophosphides. *Phys. Rev. X* **5**, 011029, doi:10.1103/PhysRevX.5.011029 (2015).

11 Xu, S.-Y. *et al.* Discovery of a Weyl fermion semimetal and topological Fermi arcs. *Science* **349**, 613-617, doi:10.1126/science.aaa9297 (2015).

12 Lv, B. Q. *et al.* Experimental Discovery of Weyl Semimetal TaAs. *Phys. Rev. X* **5**, 031013, doi:10.1103/PhysRevX.5.031013 (2015).

13 Yang, L. X. *et al.* Weyl semimetal phase in the non-centrosymmetric compound TaAs. *Nat. Phys.* **11**, 728-732, doi:10.1038/nphys3425 (2015).

14  Xu, D. F. *et al.* Observation of Fermi Arcs in non-Centrosymmetric Weyl Semi-metal Candidate NbP. *Chin. Phys. Lett.* **32**, 107101, doi:10.1088/0256-307X/32/10/107101 (2015).

15  Zhang, C. *et al.* Tantalum Monoarsenide: an Exotic Compensated Semimetal. *arXiv:1502.00251* (2015).

16  Huang, X. *et al.* Observation of the chiral-anomaly-induced negative magnetoresistance in 3D Weyl semimetal TaAs. *Phys. Rev. X* **5**, 031023, doi:10.1103/PhysRevX.5.031023 (2015).

17  Shekhar, C. *et al.* Large and unsaturated negative magnetoresistance induced by the chiral anomaly in the Weyl semimetal TaP. *arXiv:1506.06577* (2015).

18  Nielsen, H. B. & Ninomiya, M. The Adler-Bell-Jackiw anomaly and Weyl fermions in a crystal. *Phys. Lett. B* **130**, 389-396, doi:10.1016/0370-2693(83)91529-0 (1983).

19  Aji, V. Adler-Bell-Jackiw anomaly in Weyl semimetals: Application to pyrochlore iridates. *Phys. Rev. B* **85**, 241101(R), doi:10.1103/PhysRevB.85.241101 (2012).

20  Son, D. T. & Spivak, B. Z. Chiral anomaly and classical negative magnetoresistance of Weyl metals. *Phys. Rev. B* **88**, 104412, doi:10.1103/PhysRevB.88.104412 (2013).

21  Kim, H.-J. *et al.* Dirac versus Weyl Fermions in Topological Insulators: Adler-Bell-Jackiw Anomaly in Transport Phenomena. *Phys. Rev. Lett.* **111**, 246603, doi:10.1103/PhysRevLett.111.246603 (2013).

22  Hosur, P. & Qi, X. Recent developments in transport phenomena in Weyl semimetals. *C. R. Phys.* **14**, 857-870, doi:10.1016/j.crhy.2013.10.010 (2013).

23  Lv, B. Q. *et al.* Observation of Weyl nodes in TaAs. *Nat. Phys.* **11**, 724-727, doi:10.1038/nphys3426 (2015).

24  Zhang, J. L. *et al.* Pressure-induced superconductivity in topological parent compound $Bi_2Te_3$. *Proc. Natl. Acad. Sci. USA* **108**, 24-28, doi:10.1073/pnas.1014085108 (2011).

25  Zhang, C. *et al.* Phase diagram of a pressure-induced superconducting state and its relation to the Hall coefficient of $Bi_2Te_3$ single crystals. *Phys. Rev. B* **83**, 140504(R), doi:10.1103/PhysRevB.83.140504 (2011).

26  Kirshenbaum, K. *et al.* Pressure-Induced Unconventional Superconducting Phase in the Topological Insulator $Bi_2Se_3$. *Phys. Rev. Lett.* **111**, 087001, doi:10.1103/PhysRevLett.111.087001 (2013).

27  Pickard, C. J. & Needs, R. J. High-Pressure Phases of Silane. *Phys. Rev. Lett.* **97**, 045504, doi:10.1103/PhysRevLett.97.045504 (2006).

28  Pickard, C. J. & Needs, R. J. *Ab initio* random structure searching. *J Phys-Condens Mat* **23**, 053201, doi:10.1088/0953-8984/23/5/053201 (2011).

29  Kresse, G. & Furthmüller, J. Efficiency of ab-initio total energy calculations for metals and semiconductors using a plane-wave basis set. *Comput. Mater. Sci.* **6**, 15-50, doi:10.1016/0927-0256(96)00008-0 (1996).

30  Perdew, J. P., Burke, K. & Ernzerhof, M. Generalized Gradient Approximation Made Simple. *Phys. Rev. Lett.* **77**, 3865-3868, doi:10.1103/PhysRevLett.77.3865 (1996).

31  Togo, A., Oba, F. & Tanaka, I. First-principles calculations of the ferroelastic transition between rutile-type and $CaCl_2$-type $SiO_2$ at high pressures. *Phys. Rev. B* **78**, 134106, doi:10.1103/PhysRevB.78.134106 (2008).

32  Blaha, P., Schwarz, K., Madsen, G., Kvasicka, D. & Luitz, J. *WIEN2k, An Augmented Plane Wave Plus Local Orbitals Program for Calculating Crystal Properties*. (2001).

33  Kuneš, J., Novák, P., Schmid, R., Blaha, P. & Schwarz, K. Electronic structure of fcc Th: Spin-orbit calculation with $6p_{1/2}$ local orbital extension. *Phys. Rev. B* **64**, 153102, doi:10.1103/PhysRevB.64.153102 (2001).

34  Mao, H. K., Xu, J. & Bell, P. M. Calibration of the ruby pressure gauge to 800 kbar under quasi-hydrostatic conditions. *J. Geophys. Res.* **91**, 4673-4676, doi:10.1029/JB091iB05p04673 (1986).

35  Boller, H. & Parthé, E. The transposition structure of NbAs and of similar monophosphides and arsenides of niobium and tantalum. *Acta Cryst.* **16**, 1095-1101, doi:10.1107/S0365110X63002930 (1963).

36  Khveshchenko, D. V. Magnetic-Field-Induced Insulating Behavior in Highly Oriented Pyrolitic Graphite. *Phys. Rev. Lett.* **87**, 206401, doi:10.1103/PhysRevLett.87.206401 (2001).

37  Du, X., Tsai, S.-W., Maslov, D. L. & Hebard, A. F. Metal-insulator-like behavior in semimetallic bismuth and graphite. *Phys. Rev. Lett.* **94**, 166601, doi:10.1103/PhysRevLett.94.166601 (2005).

38  Zhang, C. *et al.* Observation of the Adler-Bell-Jackiw chiral anomaly in a Weyl semimetal. *arXiv:1503.02630* (2015).

39  He, L. P. *et al.* Pressure-induced superconductivity in the three-dimensional Dirac semimetal $Cd_3As_2$. *arXiv:1502.02509* (2015).

40  Zhang, S. *et al.* Breakdown of three-dimensional Dirac semimetal state in pressurized $Cd_3As_2$. *Phys. Rev. B* **91**, 165133, doi:10.1103/PhysRevB.91.165133 (2015).

41  Zhang, J. *et al.* Structural and Transport Properties of the Weyl Semimetal NbAs at High Pressure. *Chin. Phys. Lett.* **32**, 097102, doi:10.1088/0256-307X/32/9/097102 (2015).

42  Ali, M. N. *et al.* Large, non-saturating magnetoresistance in $WTe_2$. *Nature* **514**, 205-208, doi:10.1038/nature13763 (2014).

43   Soluyanov, A. A. *et al.* A New Type of Weyl Semimetals. *arXiv:1507.01603* (2015).

44   Pan, X.-C. *et al.* Pressure-driven dome-shaped superconductivity and electronic structural evolution in tungsten ditelluride. *Nat. Commun.* **6**, 7805, doi:10.1038/ncomms8805 (2015).

45   Kang, D. *et al.* Superconductivity emerging from a suppressed large magnetoresistant state in tungsten ditelluride. *Nat. Commun.* **6**, 7804, doi:10.1038/ncomms8804 (2015).

46   Liu, H. W. *et al.* Raman study of lattice dynamics in the Weyl semimetal TaAs. *Phys. Rev. B* **92**, 064302, doi:10.1103/PhysRevB.92.064302 (2015).

# Figures and Captions

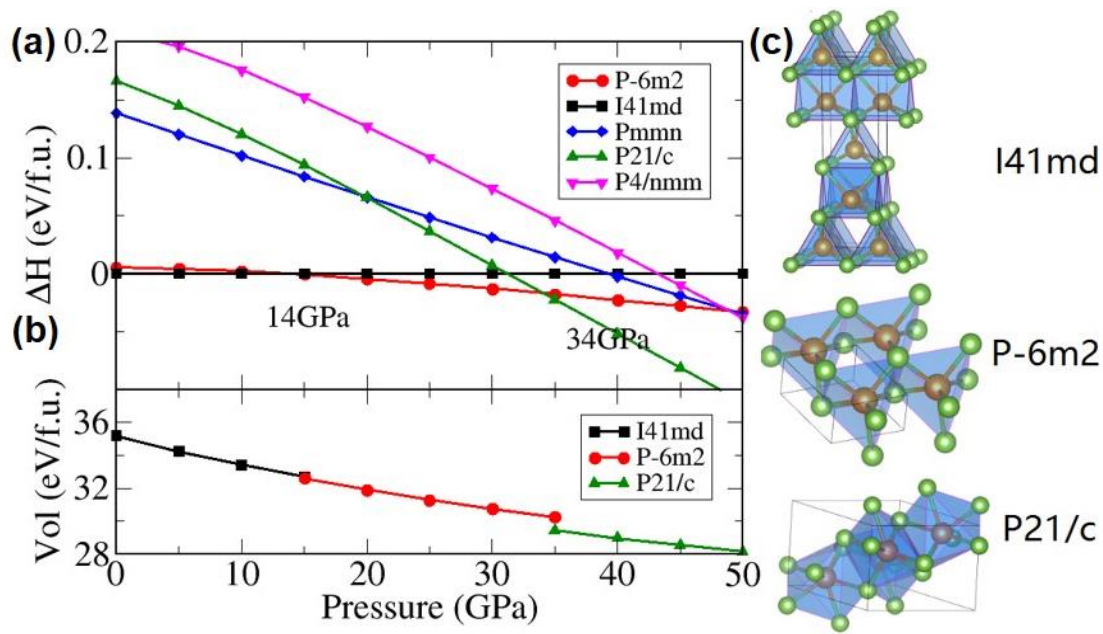

Figure 1. (a) Calculated enthalpy relative to that of *I*4$_1$*md* phase vs. pressure. (b) Volume vs. pressure. (c) Crystal structures of ambient phase (*I*4$_1$*md*) and two best candidates of high-pressure phases (*P*-6*m*2 and *P*2$_1$/*c*).

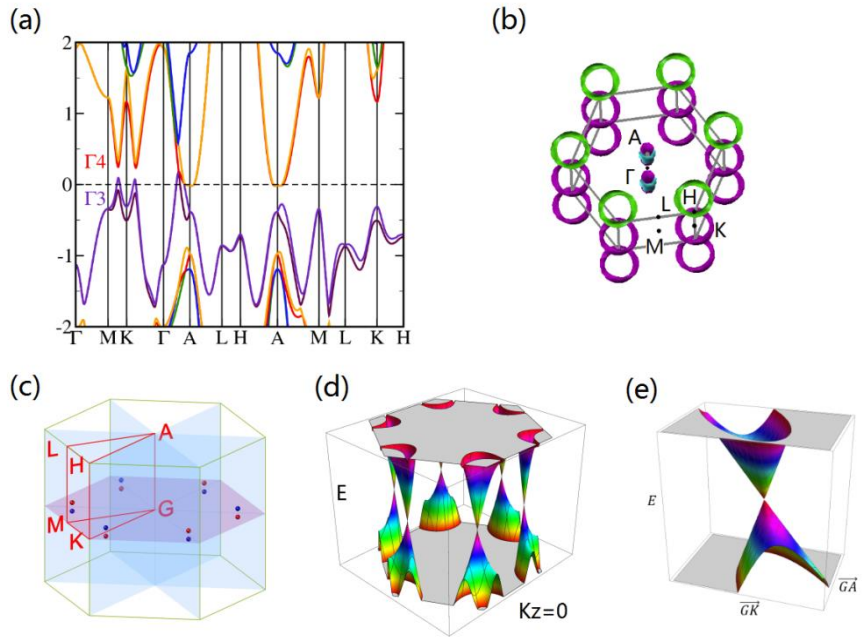

Figure 2. Calculated electronic band structures of *P*-6*m*2 phase at 25 GPa with spin-orbit coupling. (b) Fermi surfaces. (c) The Weyl nodes in the first Brillouin Zone. Weyl points obtained from scanning in the Γ-K-M plane (d) and Γ-K-A plane (e), respectively.

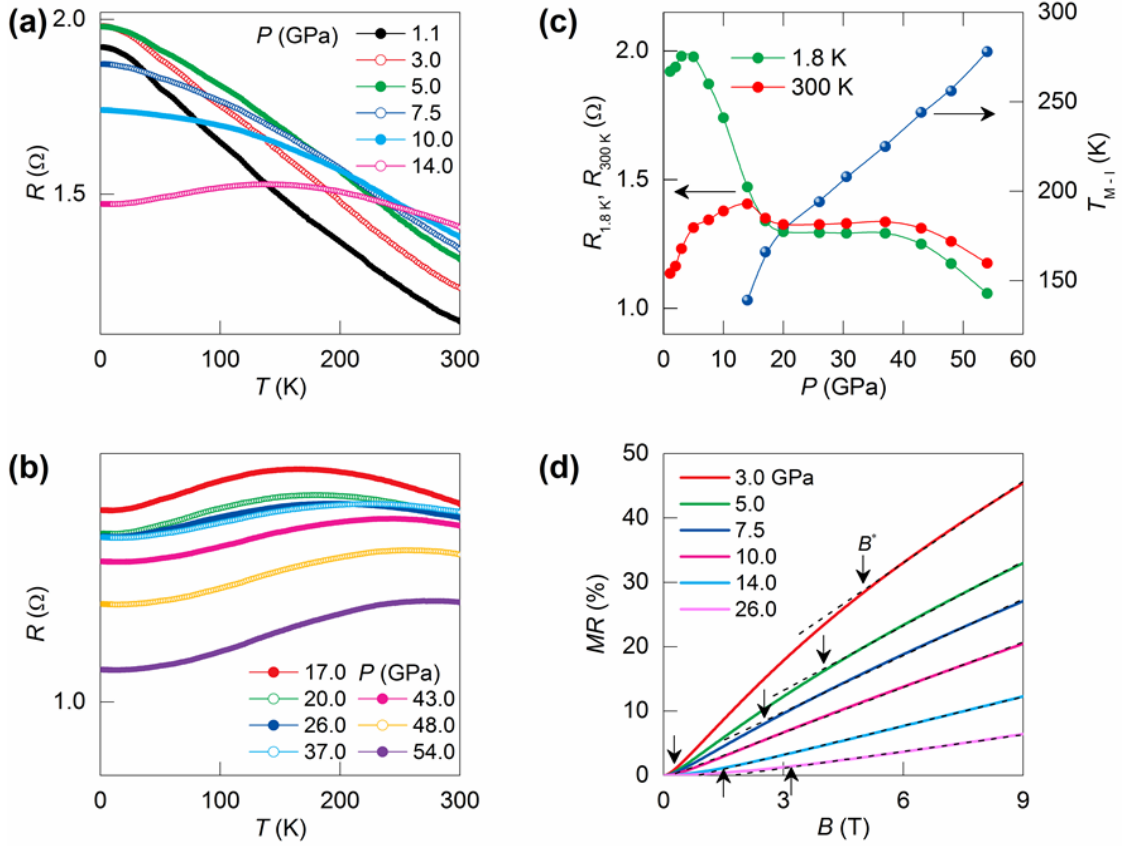

Figure 3. Temperature dependence of resistance for TaAs single crystal at various pressures. (a, b) Whole temperature resistance under high pressure from 1.1 to 54.0 GPa. (c) The insulator-metal transition temperature $T_{M-I}$ and specific resistance as a function of applied pressure at 1.8 K and 300 K, respectively. The $T_{M-I}$ is defined as the hump on the resistance curve. (d) Isothermal magnetoresistance at 4.5 K under representative pressures. The dash lines denote the linear behavior. The arrows represent the characteristic magnetic field, $B^*$, below which the magnetoresistance deviates significantly from linearity.

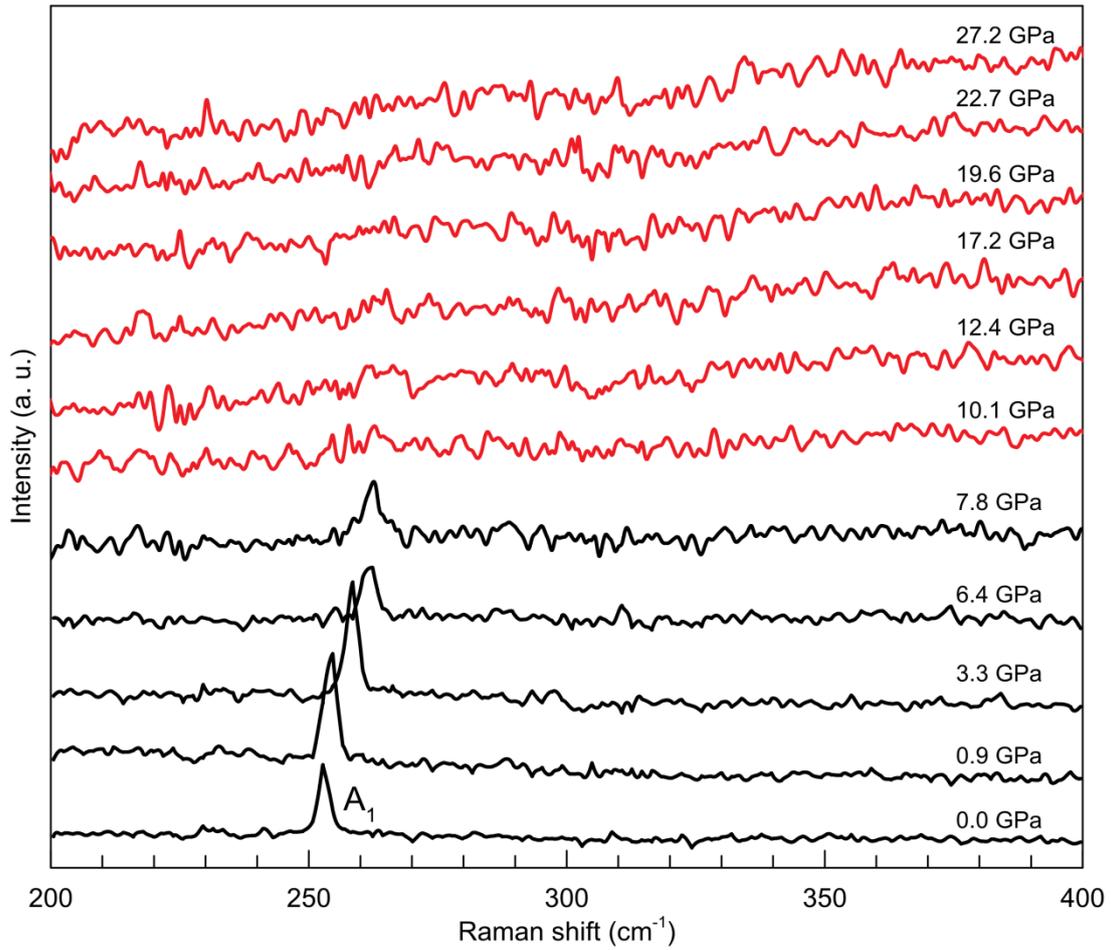

Figure 4. Raman spectra as a function of pressure at room temperature under high pressure. One peak is designated as $A_1$ mode at ambient pressure.

Table I: The coordinates of Weyl nodes of TaAs in $I4_1md$ structure. The unit of these coordinates is the same as the Ref. [10].

| Pressure (GPa) | Weyl node 1 (W1) | | | Weyl node 2 (W2) | | |
|---|---|---|---|---|---|---|
| | $x$ | $y$ | $z$ | $x$ | $y$ | $z$ |
| 0 | 0.944 | 0.0146 | 0 | 0.519 | 0.035 | 0.593 |
| 2 | 0.938 | 0.0146 | 0 | 0.521 | 0.035 | 0.587 |
| 4 | 0.933 | 0.0146 | 0 | 0.520 | 0.035 | 0.585 |
| 6 | 0.927 | 0.0146 | 0 | 0.521 | 0.035 | 0.581 |
| 8 | 0.922 | 0.0146 | 0 | 0.520 | 0.035 | 0.579 |
| 10 | 0.916 | 0.0146 | 0 | 0.520 | 0.035 | 0.576 |
| 12 | 0.910 | 0.0146 | 0 | 0.521 | 0.035 | 0.573 |
| 14 | 0.905 | 0.0144 | 0 | 0.521 | 0.035 | 0.571 |

Table II: the coordinates of Weyl nodes of TaAs in *P*-6*m*2 phase. These coordinates are almost unchanged with the increasing pressure. The position is given in the unit of the length of reciprocal lattice vectors.

| Pressure (GPa) | $x$ | $y$ | $z$ |
| --- | --- | --- | --- |
| 15 | 0.253 | 0.253 | 0.027 |
| 20 | 0.250 | 0.250 | 0.028 |
| 25 | 0.249 | 0.249 | 0.028 |
| 30 | 0.247 | 0.247 | 0.028 |

## Supplementary Information

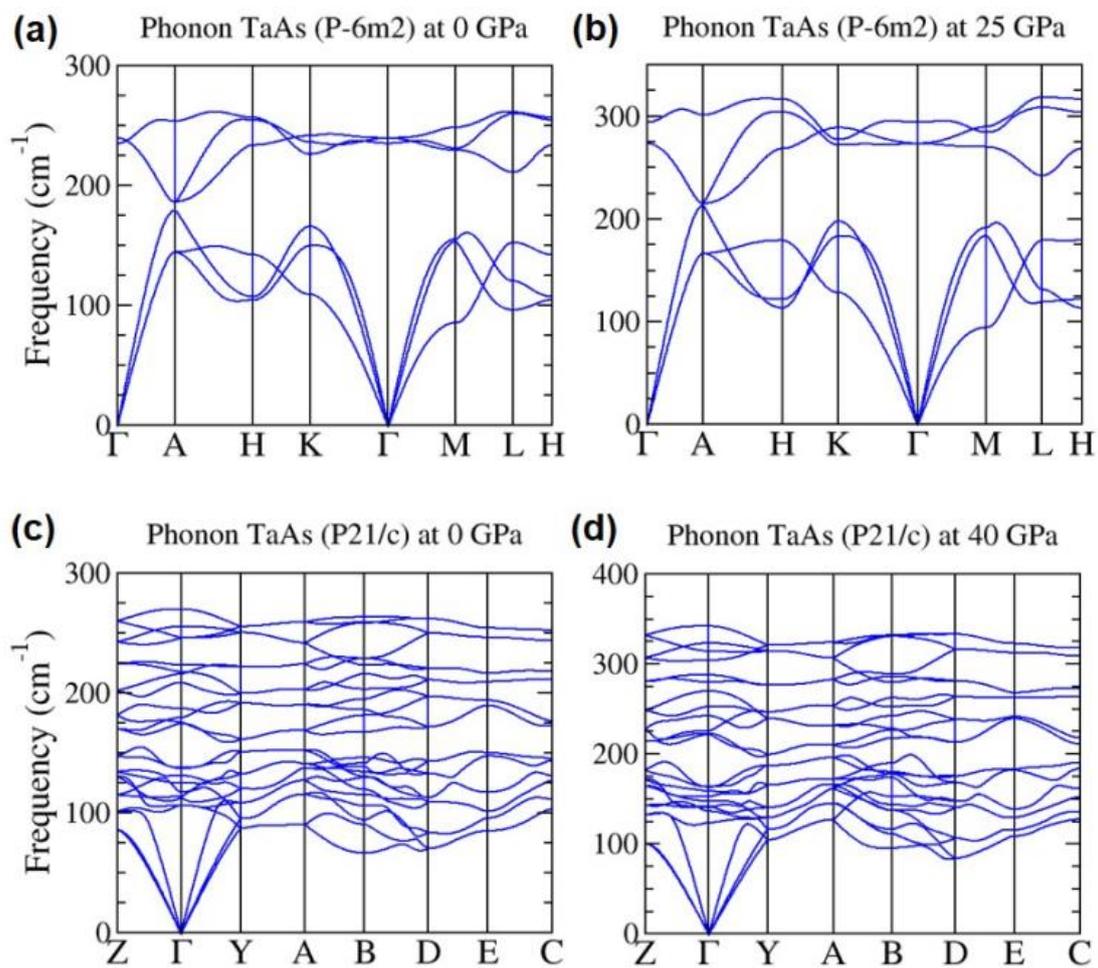

Figure S1. Phonon dispersions for the high-pressure phases of TaAs. (a) *P*-6*m*2 at 0 GPa. (b) *P*-6*m*2 at 25 GPa. (c) *P*2$_1$/*c* at 0 GPa. (d) *P*2$_1$/*c* at 40 GPa.

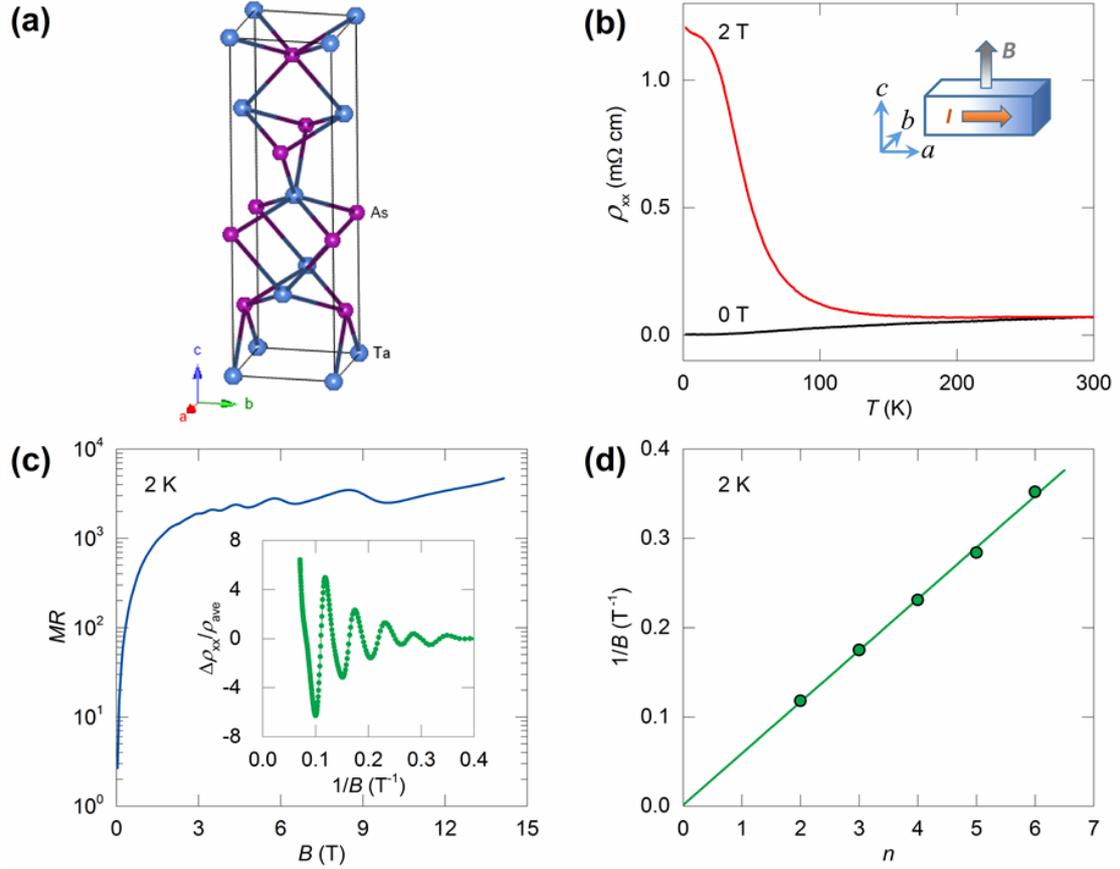

Figure S2. Schematic crystal structure and temperature-dependent in-plane resistivity for TaAs. (a) Body-centered-tetragonal NbAs-type structure of TaAs single crystal with the non-centrosymmetric space group $I4_1md$. Blue and violet balls represent the Ta atom and As atom, respectively. (b) Temperature dependence of in-plane resistivity at 0 T and 2 T, where the magnetic field is perpendicular to the *ab* plane and the current used (the inset). (c) Magnetoresistance at 2 K as a function of magnetic field perpendicular to the current. Inset: Oscillatory component of the in-plane resistivity at 2 K plotted against reciprocal magnetic field, $1/B$, obtained by subtracting a smooth background. (d) Linear fitting of the Shubnikov-de Haas fan diagram, showing the π Berry's phase of the Weyl electron pocket with a consistent intercept closest to zero.